\def\lsim {~^{<~}_{\sim~}}
\def\gsim {~^{>~}_{\sim~}}
\documentclass{PoS}

\title{SU(3) lattice QCD study of the gluon propagator \\in maximally Abelian gauge: off-diagonal gluon mass generation and infrared Abelian dominance}

\ShortTitle{SU(3) lattice QCD study of the gluon propagator in MAG}

\author{\speaker{Shinya Gongyo}\\
         Department of Physics, Kyoto University\\
        E-mail: \email{gongyo@ruby.scphys.kyoto-u.ac.jp}}

\author{Hideo Suganuma, Takumi Iritani\\
        Department of Physics, Kyoto University\\}

\abstract{We investigate gluon propagators and the effective mass of the gluon fields in the MA gauge with U(1)$_3 \times$U(1)$_8$ Landau gauge fixing in SU(3) lattice QCD. The Monte Carlo simulation is performed on $16^4$ at $\beta$=5.7, 5.8 and 6.0 and $32^4$ at $\beta=$5.8 and 6.0 at the quenched level. To calculate the propagators, we adopt a method to extract gauge fields from link-variables analytically in the SU(3) case. The off-diagonal gluons behave as massive vector bosons with the approximate effective mass $M_{\rm off} \simeq 1.1 - 1.2 {\rm GeV}$  in the region of $r = 0.3 - 0.8 {\rm fm}$, and the propagation is limited within a short range. On the other hand, the diagonal gluons behave as light vector bosons with $M_{\rm diag}\simeq 0.3 {\rm GeV} $ and the propagation of diagonal gluons remains even in a large range. In this way, infrared Abelian dominance is shown in terms of short-range propagation of off-diagonal gluons. Furthermore, we investigate the functional form of the off-diagonal gluon propagator. The functional form is well described by the four-dimensional Euclidean Yukawa-type function ${\rm exp}(-m_{\rm off} r)/r$ with $m_{\rm off} = 1.3 -1.4 {\rm GeV}$ 
for $r = 0.1- 0.8{\rm fm}$. This also indicates that the spectral function of off-diagonal gluons has the negative-value region.
}

\FullConference{Xth Quark Confinement and the Hadron Spectrum,\\
		October 8-12, 2012\\
		TUM Campus Garching, Munich, Germany}

\begin{document}

\section{Introduction}

For the quark-confinement mechanism, the dual-superconductor picture was suggested by Nambu, 't Hooft and Mandelstam \cite{N74}. In this picture, there occurs color-magnetic monopole condensation, and then the color-electric flux between the quark and the antiquark is squeezed as a one-dimensional tube due to the dual Higgs mechanism.  From the viewpoint of the dual-superconductor picture in QCD, however, there are two assumptions of Abelian dominance \cite{tH81,EI82} and monopole condensation.
Here, Abelian dominance means that only the diagonal gluon component seems to be significant to confinement.

The various lattice QCD Monte Carlo simulations  support these assumptions when the maximally abelian (MA) gauge fixing is performed \cite{KSW87,SY90,BWS91,SNW94,Mi95,Wo95,AS98,IH99,SAI02,BC03}.

According to these studies, only the diagonal gluons play a dominant role for the infrared QCD physics, which is called ``infrared Abelian dominance". Infrared Abelian dominance means that off-diagonal gluons are not significant to infrared QCD. Therefore, the essence of infrared Abelian dominance lie in the behavior of  the off-diagonal gluon propagator.

The gluon propagators in the MA gauge has been investigated in SU(2) lattice Monte Carlo simulations \cite{AS98, BC03,Cu01}.
 To investigate the gluon propagators in the MA gauge, it is desired to extract the gluons exactly from the link-variables, because the link-variable cannot be expanded even for a small lattice spacing due to large fluctuation of gluons. In SU(2) lattice case, the extraction is easy to be done without any approximation, because of the SU(2) property.
 With this extraction, the SU(2) lattice simulation suggests that the off-diagonal gluons do not propagate in the infrared region due to the effective mass $M_{\rm off} \simeq 1.2{\rm GeV}$, while the diagonal gluon widely propagates \cite{AS98}.
 
In this paper, we propose a method to extract the gluons from the link-variable directly and generally in SU(3) lattice QCD, and to investigate the gluon propagators in the MA gauge.

\section{Formalism to extract gluon fields from link-variables  }
\label{2}
In this section, we consider a useful and general method to extract the gauge fields analytically and exactly from the link-variables whether $|agA_\mu (x)| \ll 1$ is satisfied or not \cite{FN04}.

To this end, we first define the hermite matrix,
\begin{eqnarray}
	\Lambda \equiv  \frac{1}{2i}\left ( U -U^\dagger \right ) 
	= \frac{1}{2i}(e^{iagA} - e^{-iagA}) \equiv \sin agA .
\end{eqnarray}
For simplicity, we have omitted the Lorentz index and space-time arguments.

Arbitrary hermite matrix $\Lambda$ can be diagonalized by a unitary transformation as
\begin{equation}
	\Lambda _d \equiv \Omega \Lambda \Omega ^\dagger 
	=\left( 
	\begin{array}{ccc}
		\lambda_1 &  & 0 \\
		 & \lambda_2 &  \\
		0 &  & \lambda_3 \\
	\end{array} 
	\right ),
\end{equation}
where $\Omega \in$SU(3). We can obtain the eigenvalues $\lambda _i~(i=1,2,3)$ by solving 
\begin{equation}
	{\rm det}(x1 - \Lambda) = 0.
\end{equation}
This is a cubic equation on $x$ with real coefficients. The eigenvalues $\lambda _i$, i.e., the solutions of the equation are
\begin{eqnarray}
	x_{0,\pm}
	=  z_{0,\pm}\sqrt{\alpha ^2+\beta/3}+\alpha, \label{eq_cubic}
\end{eqnarray}
where 
\begin{eqnarray}
	z_0 \equiv e^{i\theta /3} +e^{-i\theta /3},~
	z_{\pm} \equiv e^{i(\theta \pm 2\pi)/3} +e^{-i(\theta \pm 2\pi)/3}, \label{eq_z}
\end{eqnarray}
and
\begin{eqnarray}
\alpha \equiv  \frac{1}{3}\mathrm{Tr}\Lambda\in {\bf R} ,
~ \beta \equiv  -(\Lambda_{22}\Lambda_{33}+\Lambda_{33}\Lambda_{11}+\Lambda_{11}\Lambda_{22}  - |\Lambda_{23}| ^2 -|\Lambda_{31}| ^2 -|\Lambda_{12}| ^2)\in {\bf R} .
\end{eqnarray}
Here, $\theta = \theta (\alpha , \beta) \in {\bf R}$ is analytically obtained and the derivation is given by Cardano's method. In this way, $\lambda_i$ is obtained. 

The unitary matrix $\Omega$ can be also derived as follows. By solving $\Lambda\vec{e}_i = \lambda _i\vec{e}_i$, we obtain normalized eigenvectors $\vec{e}_i={}^t(x_i,y_i,z_i)~(i=1,2,3)$. Because of $|\vec{e}|=1$,
there is non-zero component, so that $z_i$ is assumed to be nonzero without loss of generality and rescale it by $1/z_i$, 
\begin{equation}
	\Lambda ~{}^t\left (x_i/z_i~ y_i/z_i~ 1 \right)
	= \lambda _i
~{}^t\left (x_i/z_i~ y_i/z_i~ 1 \right).
\end{equation}
This is solved easily as
\begin{eqnarray}
	\left (
	\begin{array}{c}
		x_i /z_i \\
		y_i /z_i\\
	\end{array}
	\right )
	&=& -\{ (\Lambda_{11}-\lambda _i)(\Lambda_{22}-\lambda _i) - \Lambda_{12}\Lambda_{21} \}^{-1} 
   	\left (
		\begin{array}{cc}
		\Lambda_{22}-\lambda _i& -\Lambda_{12} \\
		-\Lambda_{21}& \Lambda_{11}-\lambda _i
	\end{array}
	\right )
	\left (
	\begin{array}{c}
		\Lambda_{13} \\
		\Lambda_{23} \\
	\end{array}
	\right ).
\end{eqnarray}
From the normalization condition $|\vec{e_i}|=1$, we obtain $\Omega ^\dagger=(\vec{e_1},\vec{e_2},\vec{e_3})$.

When we diagonalize $\Lambda$ with the unitary matrix $\Omega$, the gluon fields $A$ are diagonalized. 
Thus, the link-variables $U =e^{iagA}$ are also diagonalized with the unitary matrix $\Omega$ as 
\begin{eqnarray}
	U _d \equiv \Omega U \Omega ^\dagger
          = e^{i ag\Omega A \Omega ^\dagger } 
	\equiv
		{\rm diag}\left( 
		e^{i \theta _1} , e^{i \theta _2} , e^{i \theta _3} \right) ,
\end{eqnarray}	
where $-\pi  \leq \theta _i < \pi $ $(i=1,2,3)$ is taken. 

In this way, we can derive gluon fields $A$ from link-variables $U$ analytically by diagonalizing them, 
\begin{eqnarray}
	\Omega A \Omega ^\dagger =
			\frac{1}{ag}\left( 
	\begin{array}{ccc}
		 \theta _1 &  & 0 \\
		 &  \theta _2 &  \\
		0 &  &  \theta _3 \\
	\end{array} 
	\right)
	~\Rightarrow A = \frac{1}{ag}\Omega ^\dagger
	\left (
	\begin{array}{ccc}
		 \theta _1 &  & 0 \\
		 &  \theta _2 &  \\
		0 &  &  \theta _3 \\
	\end{array} 
	\right)
	\Omega. 
\end{eqnarray}
This formalism is quite general, because the derivation is correct in any gauge and even without any gauge fixing.

\section{SU(3) lattice  QCD results for gluon propagators in the MA gauge}
\label{3}

Using the SU(3) lattice QCD, we calculate the gluon propagators \cite{GIS12} in the MA gauge with the U(1)$_3\times$U(1)$_8$ Landau gauge fixing. In the MA gauge, to investigate the gluon propagators, we use the above-mentioned method. The Monte Carlo simulation is performed  with the standard plaquette action on the $16^4$ lattice with $\beta$ =5.7, 5.8 and 6.0 and $32^4$ with $\beta$ = 5.8 and 6.0 at the quenched level. All measurements are done every 500 sweeps after a thermalization of 10,000 sweeps using the pseudo heat-bath algorithm. We prepare 50 configurations with $16^4$ and 20 configurations with $32^4$ at each $\beta$. The statistical error is estimated with the jackknife method.

Here, we study the Euclidean scalar combination of 
the diagonal (Abelian) and off-diagonal gluon propagators as
\begin{eqnarray}
G_{\mu\mu}^{\rm Abel}(r) \equiv  \frac{1}{2} \sum_{a= 3,8} \left< A_\mu^a(x)A_\mu^a(y)\right>,~
				\label{eqn:AAf002}
G_{\mu\mu}^{\rm off}(r) \equiv 
\frac{1}{6} \sum_{a\neq 3,8} \left< A_\mu^a(x)A_\mu^a(y)\right>.
				\label{eqn:AAf003}
\end{eqnarray}
The scalar combination of the propagator is expressed as the function of the four-dimensional Euclidean distance $r\equiv \sqrt{(x_\mu -y_\mu)^2}$. 
When we consider the renormalization, 
these propagators are multiplied by an $r$-independent constant, according 
to a constant renormalization factor of the renormalized gluon fields.

\begin{figure}[h]
\begin{center}
\includegraphics[scale=0.5]{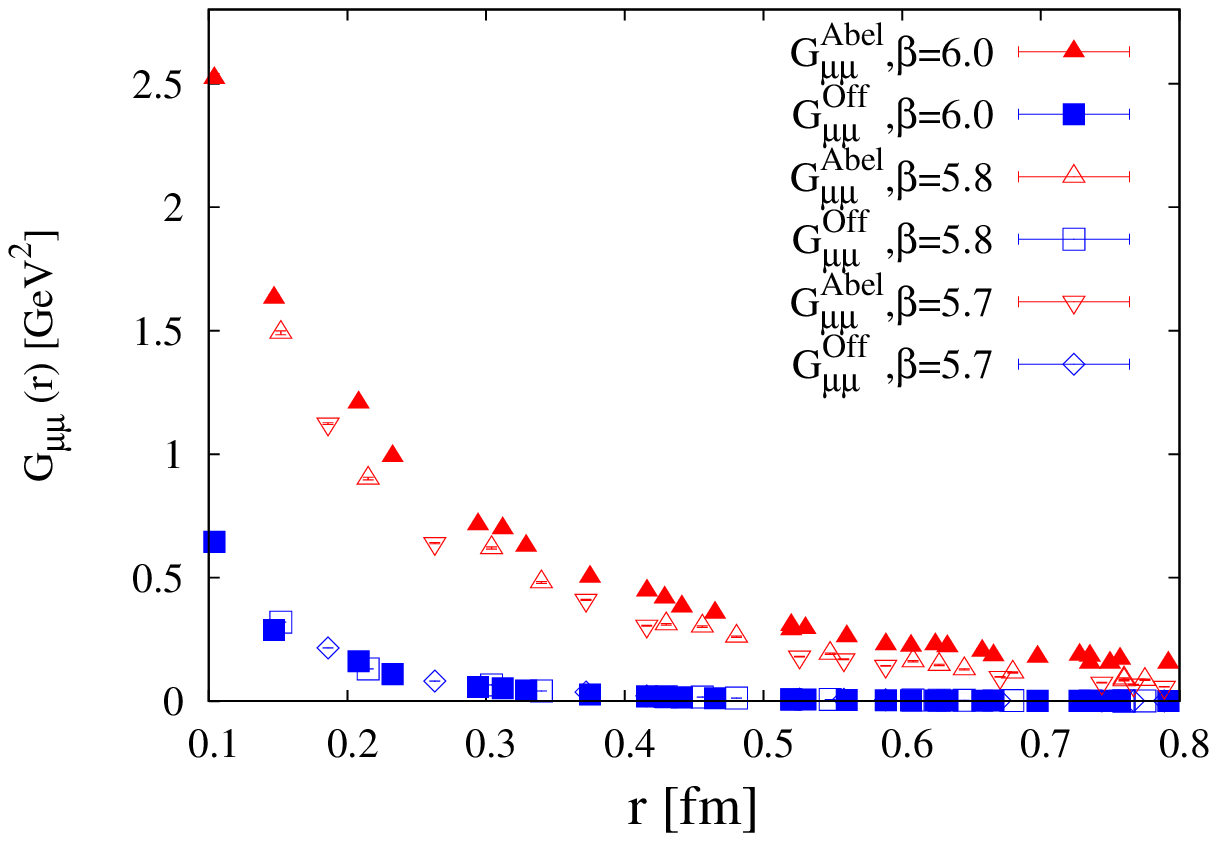}
\includegraphics[scale=0.5]{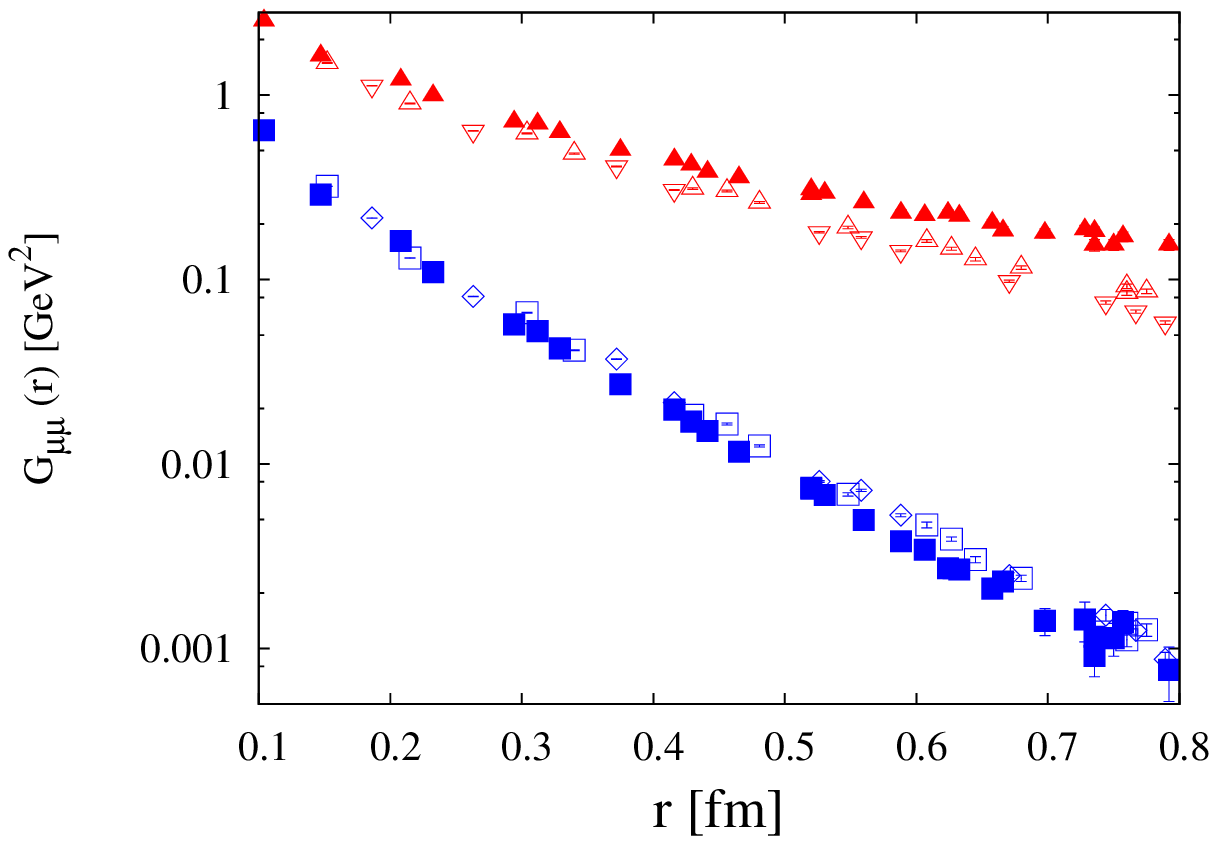}
\caption{
The SU(3) lattice QCD results of the gluon propagators $G_{\mu\mu}^{\rm Abel}(r)$ and $G_{\mu\mu}^{\rm off}(r)$ as the function of $r\equiv \sqrt{(x_\mu -y_\mu)^2}$ in the MA gauge with the U(1)$_3\times$U(1)$_8$ Landau gauge fixing in the physical unit. The Monte Carlo simulation is performed on the $16^4$ lattice with $\beta$ = 5.7, 5.8 and 6.0. The diagonal-gluon propagator $G_{\mu\mu}^{\rm Abel}(r)$ takes a large value even at the long distance. On the other hand, the off-diagonal gluon propagator $G_{\mu\mu}^{\rm off}(r)$ rapidly decreases.
}
\label{Fig1}
\end{center}
\end{figure}

We show in Fig.\ref{Fig1} the lattice QCD result for the diagonal gluon propagator $G_{\mu\mu}^{\rm Abel}(r)$ and the off-diagonal gluon propagator $G_{\mu\mu}^{\rm off}(r)$ in the MA gauge with the U(1)$_3\times$U(1)$_8$ Landau gauge fixing. In the MA gauge, $G_{\mu\mu}^{\rm Abel}(r)$ and 
$G_{\mu\mu}^{\rm off}(r)$ manifestly differ.
The diagonal-gluon propagator $G_{\mu\mu}^{\rm Abel}(r)$ 
takes a large value even at the long distance. 
In fact, the diagonal gluons $A_\mu^3,A_\mu^8$ in the MA gauge 
propagate over the long distance.
In contrast, the off-diagonal gluon propagator 
$G_{\mu\mu}^{\rm off}(r)$ rapidly decreases and is negligible
for $r \gsim 0.4$fm in comparison with $G_{\mu\mu}^{\rm Abel}(r)$. 
Then, the off-diagonal gluons $A_\mu^a~(a\neq 3,8)$ seem to propagate 
only within the short range as $r \lsim 0.4$fm.
Thus, ``infrared abelian dominance" is found in the MA gauge.

\section{Estimation of diagonal and off-diagonal gluon mass in the MA gauge}
\label{4}
Next, we investigate the effective mass of diagonal and off-diagonal gluons \cite{GIS12}.
We start from the Lagrangian of 
the free massive vector field $A_\mu$ with the mass $M \ne 0$ 
in the Euclidean metric. In the infrared region with large $Mr$, 
the propagator $G_{\mu\mu}(r;M)$ reduces to 
\begin{eqnarray}
{G}_{\mu\mu}(r;M) &=& \left< A_\mu(x) A_\mu(y) \right> 
=\frac{3}{4\pi^2}\frac{M}{r}K_1(Mr)
	\simeq
	\frac{3\sqrt{M}}{2(2\pi)^{\frac{3}{2}}} \frac{e^{-Mr}}{r^\frac{3}{2}}, 
						\label{eqn:prp03} 
\end{eqnarray}
\begin{figure}[h]
\begin{center}
\includegraphics[scale=0.4]{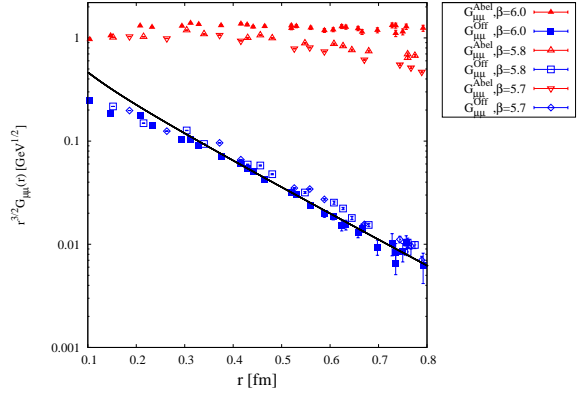}
\caption{
The logarithmic plot of $r^{3/2}G_{\mu\mu}^{\rm off} (r)$ and $r^{3/2}G_{\mu\mu}^{{\rm Abel}} (r)$ as the function of the Euclidean distance $r$ in the MA gauge with the U(1)$_3\times$U(1)$_8$ Landau gauge fixing, in the SU(3) lattice QCD with $16^4$ at $\beta$ = 5.7, 5.8 and 6.0. The solid line denotes the logarithmic plot of $r^{3/2}G_{\mu\mu}(r) \sim r^{1/2}K_1 (Mr)$ in the Proca formalism.
}
\label{Fig2}
\end{center}
\end{figure}

\begin{figure}[h]
\begin{center}
\includegraphics[scale=0.4]{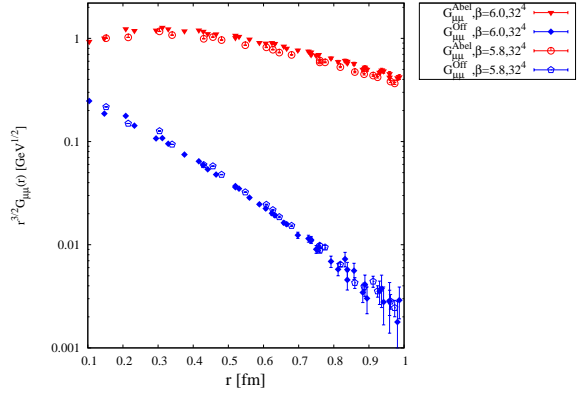}
\caption{
The logarithmic plot of $r^{3/2}G_{\mu\mu}^{\rm off} (r)$ and $r^{3/2}G_{\mu\mu}^{{\rm Abel}} (r)$ as the function of the Euclidean distance $r$ in the MA gauge with the U(1)$_3\times$U(1)$_8$ Landau gauge fixing, in the SU(3) lattice QCD with $32^4$ at $\beta$ = 5.8 and 6.0.
}
\label{Fig2-2}
\end{center}
\end{figure}

In Fig.\ref{Fig2}, we show the logarithmic plot of $r^{3/2}G_{\mu\mu}^{\rm off} (r)$ and $r^{3/2}G_{\mu\mu}^{\rm Abel} (r)$ as the function of the Euclidean distance $r$ in the MA gauge with the U(1)$_3\times$U(1)$_8$ Landau gauge fixing. From the linear slope on $r^{3/2}G_{\mu\mu}^{\rm off} (r)$ in the range of $r=0.3-0.8~{\rm fm}$, the effective off-diagonal gluon mass $M_{\rm off}$ is estimated. 
Note that the gluon-field renormalization does not affect the gluon mass 
estimate, since it gives only an overall constant factor for the propagator. 
We summarize in Table~1 the effective off-diagonal gluon mass $M_{\rm off}$ 
obtained from the slope analysis 
 with $16^4$ at $\beta$ =5.7, 5.8 and 6.0.
Therefore, the off-diagonal gluons seem to have a large mass
$M_{\rm off} \simeq 1.1-1.2~{\rm GeV}$.
This result approximately coincides with SU(2) lattice calculation \cite{AS98}.

On the other hand, for the diagonal gluons, their propagator seems to have some dependence on $\beta$ and lattice volume. Therefore, we estimate the effective diagonal gluon mass $M_{\rm diag}$ with larger lattice size, $32^4$. In Fig.\ref{Fig2-2}, we present the logarithmic plot of $r^{3/2}G_{\mu\mu}^{\rm off} (r)$ and $r^{3/2}G_{\mu\mu}^{{\rm Abel}} (r)$ with $32^4$ at $\beta=$5.8 and 6.0. From the linear slope on $r^{3/2}G_{\mu\mu}^{\rm Abel} (r)$ in the range of $r=0.3-0.8~{\rm fm}$, the effective diagonal gluon mass is estimated as $M_{\rm diag} \simeq 0.3 {\rm GeV}$ at each $\beta$. 

\begin{table}[h]
\caption{Summary table of conditions and results in SU(3) lattice QCD. In the MA gauge, the off-diagonal gluons seem to have a large effective mass $M_{\rm off} \simeq 1.1-1.2~\mathrm{GeV}$ and the functional form in the range of $r=0.1-0.8~{\rm fm}$ is well described with the four-dimensional Euclidean Yukawa function $\sim \exp(-m_{\rm off}r)/r$  with $m_{\rm off} \simeq 1.3-1.4~\mathrm{GeV}$. }
\begin{center} 
\begin{tabular}{ccccc}
\hline \hline
  lattice size   & $\beta$     & $a[{\rm fm}]$ &   $M_{\rm off} [{\rm GeV}] $  & $m_{\rm off} [{\rm GeV}] $ \\
\hline
                 &    5.7  & 0.186 &  1.2  & 1.3 \\
$16^4$                 &    5.8  & 0.152 &  1.1 & 1.3 \\
                 &    6.0     &  0.104 & 1.1 & 1.4 \\
\hline \hline
\end{tabular}
\end{center} 
\end{table}
Finally in this section, we discuss the relation 
between infrared abelian dominance and 
the off-diagonal gluon mass.
Due to the large effective mass $M_{\rm off} $, 
the off-diagonal gluon propagation is restricted within about
$M_{\rm off}^{-1} \simeq 0.2$fm in the MA gauge.
Therefore, at the infrared scale as $r \gg 0.2{\rm fm}$,
the off-diagonal gluons $A_\mu^a~(a\neq 3,8)$ cannot mediate the long-range force like the massive weak bosons in the Weinberg-Salam model, 
and only the diagonal gluons $A_\mu^3,~A_\mu^8$ can mediate  
the long-range interaction in the MA gauge.
In fact, in the MA gauge, the off-diagonal gluons are expected to be 
inactive due to the large mass $M_{\rm off}$ in the infrared region 
in comparison with the diagonal gluons. 
Then, infrared abelian dominance holds for $r \gg M^{-1}_{\rm off}$. 

\section{Analysis of the functional form of off-diagonal gluon propagator in the MA gauge}
\label{4}
In this section, we investigate the functional form of the off-diagonal gluon propagator $G_{\mu\mu}^{\rm off}(r)$ in the MA gauge in SU(3) lattice QCD \cite{GIS12}. In the previous section, we compare the off-diagonal gluon propagator with the massive vector boson propagator and estimate the gluon mass. In fact, the gluon propagator would not be described by a simple massive propagator in the whole region of $r=0.1-0.8~{\rm fm}$.

There is the similar situation in the Landau gauge \cite{IS09}. The functional form of the gluon propagator cannot be described by $\exp(-Mr)/r^{3/2}$ with an effective mass $M$ in the whole region of $r=0.1-1.0~{\rm fm}$. The appropriate form is the four-dimensional Euclidean Yukawa-type function $\exp(-mr)/r$ with a mass parameter $m$.

\begin{figure}[h]
\begin{center}
\includegraphics[scale=0.5]{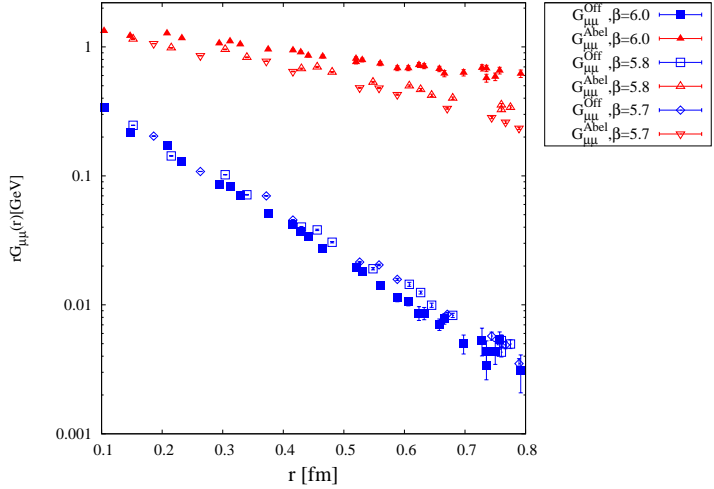}
\caption{
The logarithmic plot of $rG_{\mu\mu}^{\rm off} (r)$ and $rG_{\mu\mu}^{{\rm Abel}} (r)$ as the function of the Euclidean distance $r$ in the MA gauge with the U(1)$_3\times$U(1)$_8$ Landau gauge fixing, using the SU(3) lattice QCD with $16^4$ at $\beta$=5.7, 5.8 and 6.0. For $rG_{\mu\mu}^{\rm off} (r)$, the approximate linear correlaton is found.}
\label{Fig3}
\end{center}
\end{figure}

In the same way, in the MA gauge, we also compare the gluon propagator with the four-dimensional Euclidean Yukawa function.
 In Fig.\ref{Fig3}, we show the logarithmic plot of $rG_{\mu\mu}^{\rm off} (r)$ and $rG_{\mu\mu}^{\rm Abel} (r)$ as the function of the distance $r$ in the MA gauge with the U(1)$_3\times$U(1)$_8$ Landau gauge fixing.
Note that the logarithmic plot of $rG_{\mu\mu}^{\rm off} (r)$ is almost linear in the whole region of $r=0.1-0.8~{\rm fm}$, 
and therefore the off-diagonal gluon propagator is well expressed by the four-dimensional Euclidean Yukawa function $Ae^{-m_{\rm off} r}/r$, 
with a mass parameter $m_{\rm off}$ and a dimensionless constant $A$.
The best-fit mass parameter $m_{\rm off}$ is given in Table 1 at each $\beta$ = 5.7, 5.8 and 6.0. 

We comment on the four-dimensional Euclidean Yukawa-type propagator \cite{IS09}. If the functional form of the off-diagonal gluon is well described by the four-dimensional Yukawa function, we analytically calculate the off-diagonal zero-spatial-momentum propagator,
 $D_0^{\rm off}  (t)\equiv \int d^3x G_{\mu\mu}^{\rm off} (r)$, and obtain 
the spectral function $\rho^{\rm off} (\omega)$ by the inverse Laplace transformation in the MA gauge: 
\begin{eqnarray}
\rho^{\rm off} (\omega) = - \frac{4\pi Am_{\rm off}}{(\omega ^2 -m_{\rm off}^2)^{3/2}}\theta (\omega -m_{\rm off}) + \frac{4\pi A/\sqrt{2m_{\rm off}}}{(\omega - m_{\rm off})^{1/2}}\delta(\omega -m_{\rm off}).
\end{eqnarray} 

\section{Summary and Concluding Remarks}
\label{5}
We have performed the study of the gluon propagators in the MA gauge 
with the U(1)$_3\times$U(1)$_8$ Landau gauge fixing in the SU(3) quenched 
lattice QCD. 
 To investigate the gluon propagators in the MA gauge, we have considered to derive the gluon fields analytically from the SU(3) link-variables.
 
With this method, we have calculated the Euclidean scalar combination 
$G_{\mu\mu} (r)$ of the diagonal and the off-diagonal gluon propagators, 
and have considered the origin of infrared Abelian dominance. 
The Monte Carlo simulation is performed on the $16^4$ lattice at 
$\beta$=5.7, 5.8 and 6.0 and on the $32^4$ at $\beta$=5.8 and 6.0 at the quenched level.
We have found that 
the off-diagonal gluons behave as massive vector bosons 
with the effective mass $M_{\rm off} \simeq 1.1-1.2$~GeV for $r =0.3-0.8$~fm. 
The effective gluon mass has been estimated from the linear fit analysis of the logarithmic plot of $r^{3/2}G_{\mu\mu} ^{\rm off}(r)$. 
Due to the large value, the finite-size effect for the off-diagonal gluon mass 
is expected to be ignored. The large gluon mass shows that the off-diagonal 
gluons cannot mediate the interaction over the large distance as 
$r \gg M_{\rm off}^{-1}$, and such an infrared inactivity of the off-diagonal 
gluons would lead infrared Abelian dominance in the MA gauge.

On the other hand, from the behavior of the diagonal gluon propagator 
$G_{\mu\mu} ^{\rm Abel}(r)$ and $r^{3/2}G_{\mu\mu} ^{\rm Abel}(r)$, 
 the diagonal gluons seem to behave as light vector bosons with $M_{\rm diag} \simeq 0.3$~GeV for $r =0.3-0.8$~fm \cite{GIS12}, considering also the larger-volume analysis with $32^4$ at $\beta=5.8$ and $6.0$.

Finally, we have also investigated the functional form of the off-diagonal 
gluon propagator $G_{\mu\mu} ^{\rm off}(r)$ in the MA gauge. 
We show that $G_{\mu\mu} ^{\rm off}(r)$ is well described by the 
four-dimensional Euclidean Yukawa-type form with the mass parameter 
$m_{\rm off} \simeq 1.3 -1.4$~GeV in the whole region of $r=0.1-0.8$~fm.
This indicates that the spectral function $\rho^{\rm off} (\omega)$ of 
the off-diagonal gluons in the MA gauge 
has the negative-value region \cite{GIS12}, 
as in the Landau gauge \cite{IS09,MO87,Bo04}.

%On the other hand, the functional form of the diagonal gluon propagator 
%seems to be the four-dimensional Euclidean Yukawa function with the lighter 
%mass parameter. However, to discuss the functional form clearly, the finite 
%size effect is to be checked carefully just like the estimation of the 
%diagonal effective gluon mass. 

\section*{Acknowledgements}
This work is supported in part by the Grant for Scientific Research 
[(C) No.~23540306, Priority Areas ``New Hadrons'' (E01:21105006)], Grant-in-Aid for JSPS Fellows (No.23-752, 24-1458)
from the Ministry of Education, Culture, Science and Technology 
(MEXT) of Japan, and the Global COE Program at Kyoto University.
The lattice QCD calculations are done on NEC SX-8R at Osaka University.


\begin{thebibliography}{99}
\bibitem{N74}
Y.~Nambu, {\it Phys. Rev.} {\bf D10} (1974) 4262; 
G.~'t Hooft, in {\it High Energy Physics} (1975); \\
S.~Mandelstam, {\it Phys. Rept.} {\bf 23} (1976) 245 .

\bibitem{tH81}	G.~'t Hooft, {\it Nucl. Phys.} {\bf B190} (1981) 455.

\bibitem{EI82}
Z.F.~Ezawa and A.~Iwazaki, {\it Phys. Rev.} {\bf D25} (1982) 2681;
{\it Phys. Rev.} {\bf D26} (1982) 631.

\bibitem{KSW87} A.~S.~Kronfeld, G.~Schierholz and U.-J.~Wiese, 
{\it Nucl. Phys.} {\bf B293} (1987) 461; \\
A.~S.~Kronfeld, M.~L.~Laursen, G.~Schierholz and U.-J.~Wiese, 
{\it Phys. Lett.} {\bf B198} (1987) 516.

\bibitem{SY90}	T. Suzuki and I. Yotsuyanagi, 
{\it Phys. Rev.} {\bf D42} (1990) 4257.

\bibitem{BWS91} F. Brandstaeter, U.-J. Wiese and G. Schierholz, 
{\it Phys. Lett.} {\bf B272} (1991) 319.

\bibitem{SNW94} J.~D.~Stack, S.~D.~Neiman and R.~J.~Wensley, 
{\it Phys. Rev.}  {\bf D50} (1994) 3399.

\bibitem{Mi95} O.~Miyamura, {\it Phys. Lett.} {\bf B353} (1995) 91.

\bibitem{Wo95} R.~M.~Woloshyn, {\it Phys. Rev.} {\bf D51} (1995) 6411.

\bibitem{AS98}
K.~Amemiya and H.~Suganuma, {\it Phys. Rev.} {\bf D60} (1999) 114509.

\bibitem{IH99} H.~Ichie and H.~Suganuma, 
{\it Nucl. Phys.} {\bf B548} (1999) 365; 
{\it Nucl. Phys.} {\bf B574} (2000) 70.

\bibitem{SAI02}
H.~Suganuma et al.,
%K.~Amemiya, H.~Ichie, N.~Ishii, H.~Matsufuru and T.~T.~Takahashi, 
{\it Nucl. Phys. Proc. Suppl.} {\bf 106} (2002) 679;
A.~Shibata et al., {\it PoS} (Lattice2007) 331.

\bibitem{BC03}
V.~G.~Bornyakov, M.~N.~Chernodub, F.~V.~Gubarev, S.~M.~Morozov 
and M.~I.~Polikarpov, \\
{\it Phys. Lett.} {\bf B559} (2003) 214.

\bibitem{Cu01}
A.~Cucchieri, F.~Karsch and P.~Petreczky, 
{\it Phys. Lett.} {\bf B497} (2001) 80.

\bibitem{FN04}
S.~Furui and H.~Nakajima, {\it Phys. Rev.} {\bf D69} (2004) 074505; 
{\it Nucl. Phys. Proc. Suppl.} {\bf 73} (1999) 865.

\bibitem{GIS12}
S. Gongyo, T. Iritani and H. Suganuma, 
{\it Phys. Rev.} {\bf D86} (2012) 094018;
{\it PoS} (Lattice2012) 212.

\bibitem{IS09}
T.~Iritani, H.~Suganuma and H.~Iida, {\it Phys. Rev.} {\bf D80} (2009) 114505.

\bibitem{MO87}
J.~E.~Mandula and M.~Ogilvie, {\it Phys. Lett.} {\bf B185} (1987) 127.

\bibitem{Bo04}
 P.~O.~Bowman et al., {\it Phys. Rev.} {\bf D70} (2004) 034509;
{\it Phys. Rev.} {\bf D76} (2007) 094505.

\end{thebibliography}
\end{document}